\documentclass[aps,prl,lengthcheck,superscriptaddress,amsmath,amssymb,longbibliography]{revtex4-1}
\usepackage[utf8]{inputenc}
\usepackage[T1]{fontenc}

\usepackage{graphicx}
\usepackage{bm} % bold math
\usepackage{mathbbol} % bold math
\usepackage[table,usenames,dvipsnames]{xcolor}
\usepackage[colorlinks=true,allcolors=blue]{hyperref}

\usepackage{microtype}

\def\equationautorefname~#1\null{Eq.~(#1)\null}

\renewcommand{\Im}{\operatorname{Im}}

\newcommand{\Jmod}{J_{\mathrm{mod}}}
\newcommand{\Heff}{\tilde{\mathbf{H}}}

\begin{document}

\title{Few-mode Field Quantization of Arbitrary Electromagnetic Spectral Densities}

\author{Ivan Medina}
\affiliation{Departamento de F\'isica Te\'orica de la Materia
Condensada and Condensed Matter Physics Center (IFIMAC),
Universidad Aut\'onoma de Madrid, E-28049 Madrid, Spain}
\affiliation{Centro de Ci\^encias Naturais e Humanas, Universidade
Federal do ABC, 09210-170, Santo Andr\'e, Sao P\~aulo, Brazil}

\author{Francisco J. Garc\'ia-Vidal}
\affiliation{Departamento de F\'isica Te\'orica de la Materia
Condensada and Condensed Matter Physics Center (IFIMAC),
Universidad Aut\'onoma de Madrid, E-28049 Madrid, Spain}
\affiliation{Donostia International Physics Center (DIPC), E-20018
Donostia/San Sebasti\'an, Spain}

\author{Antonio I. Fern\'andez-Dom\'inguez}
\email{a.fernandez-dominguez@uam.es} \affiliation{Departamento de
F\'isica Te\'orica de la Materia Condensada and Condensed Matter
Physics Center (IFIMAC), Universidad Aut\'onoma de Madrid, E-28049
Madrid, Spain}

\author{Johannes Feist}
\email{johannes.feist@uam.es} \affiliation{Departamento de
F\'isica Te\'orica de la Materia Condensada and Condensed Matter
Physics Center (IFIMAC), Universidad Aut\'onoma de Madrid, E-28049
Madrid, Spain}

\begin{abstract}
We develop a framework that provides a few-mode master equation
description of the interaction between a single quantum emitter
and an arbitrary electromagnetic environment. The field
quantization requires only the fitting of the spectral density,
obtained through classical electromagnetic simulations, to a model
system involving a small number of lossy and interacting modes. We
illustrate the power and validity of our approach by describing
the population and electric field dynamics in the spontaneous
decay of an emitter placed in a complex hybrid plasmonic-photonic
structure.
\end{abstract}

\maketitle

%\section{Introduction}

Over the last years, there has been large interest in developing
strategies for quantizing electromagnetic (EM) modes in open,
dispersive and absorbing photonic environments. This has been
motivated largely by the desire to use nanophotonic devices for
quantum optics and quantum technology applications. This is a
theoretical challenge, as standard ways of obtaining quantized
modes are not valid in these systems~\cite{Koenderink2010,
Kristensen2012}. In principle, macroscopic quantum electrodynamics
(QED) is the framework for such quantization in material
structures described by EM constitutive relations~\cite{Fano1956,
Huttner1992, Scheel1998, Knoll2001, Scheel2008, Buhmann2012I, Buhmann2012II}.
However, in this formalism, electromagnetic fields are
described by a continuum of harmonic oscillators, restricting its applicability
to cases where they can be treated perturbatively or
eliminated by Laplace transform or similar techniques. Work on
specific structures has focused on (possibly approximately)
obtaining quantized few-mode descriptions for plasmonic geometries
such as a metal surface~\cite{Gonzalez-Tudela2014}, metallic
spheres~\cite{Waks2010, Delga2014, Rousseaux2016,
Varguet2019Non-hermitian} or sphere
dimers~\cite{Li2016Transformation, Cuartero-Gonzalez2018}. It is
thus desirable to develop tractable but versatile models using only a small number of EM modes. During
the past few decades, there has been extensive work in this
direction, with one notable development given by pseudomode
theory~\cite{Imamoglu1994, Garraway1997Decay,
*Garraway1997Nonperturbative, Dalton2001}.

Within nanophotonics increasing attention
has recently focused on hybrid metallodielectric
setups~\cite{Doeleman2016,Peng2017,Gurlek2018,Franke2019}, with
the objective of combining the strong field confinement and
enhanced light-matter interactions of plasmonic resonances with
the long-lived nature (large quality factors) of microcavity or
photonic crystal modes. In these systems, EM field quantization is
particularly complex due to the inherent coexistence of modes with
very different properties and their mutual coupling. Quasinormal
modes~\cite{Ching1998,Kristensen2014,Gurlek2018,Lalanne2018} can
be useful to unveil the EM mode structure, but due to their lossy
nature, direct quantization remains challenging, and has only been
carried out within very limited spectral
windows~\cite{Hughes2018,Franke2019}. A complementary technique
developed very recently in the context of X-ray quantum optics is based on a
partition of the physical space~\cite{Lentrodt2020}.

\begin{figure}[t]
\includegraphics[width=\linewidth]{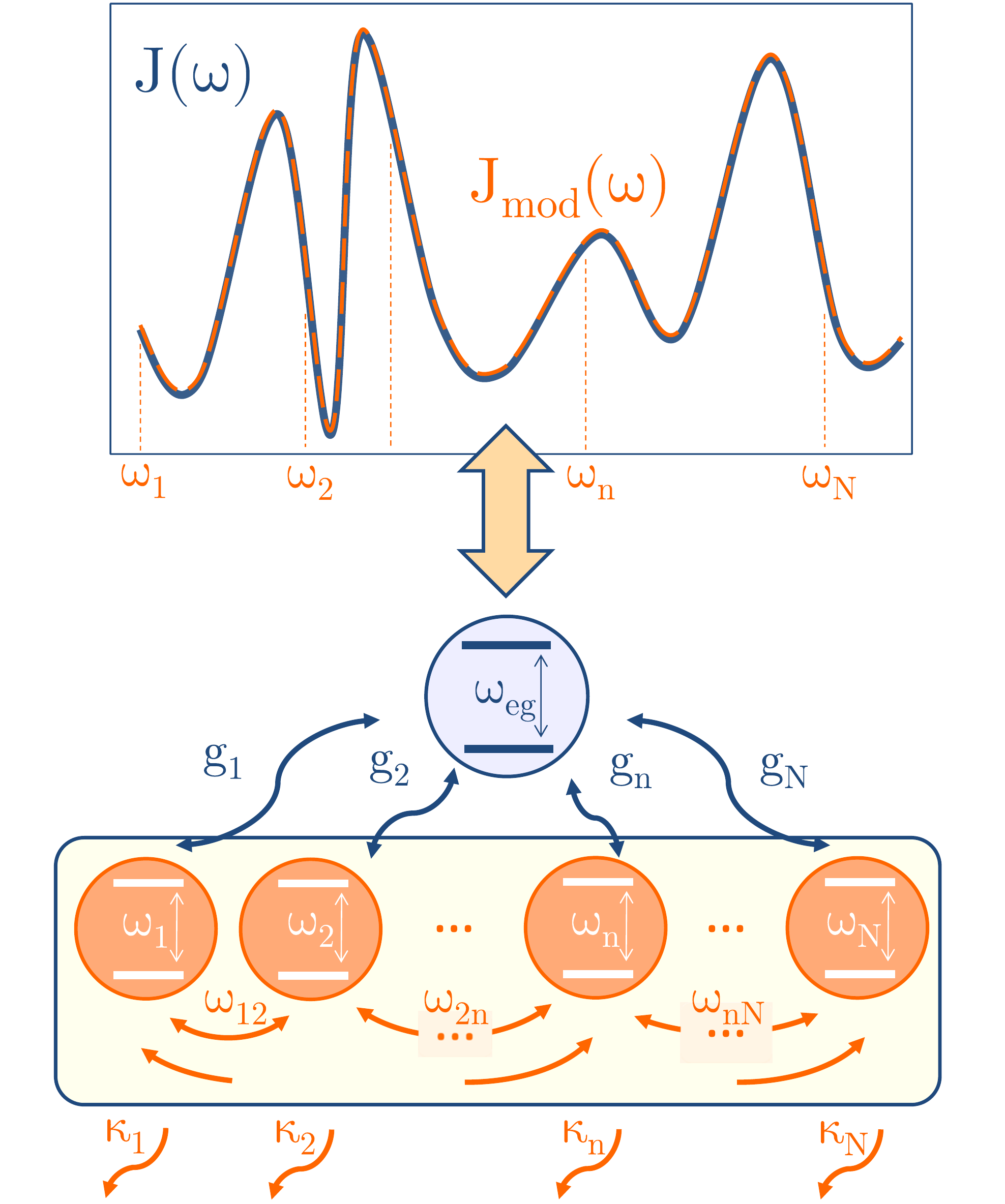}
\caption{Sketch of the model and quantization approach. The
fitting $\Jmod(\omega)$, given by \autoref{eq:Jmodel}, of an
original spectral density, $J(\omega)$ (obtained usually from a
numerical solution of Maxwell's equations), provides a natural
few-mode EM field quantization embodied by the master equation in
\autoref{eq:Lmodel}.} \label{fig:1}
\end{figure}

It is well-known that the interaction of a single emitter with an
arbitrary EM environment can be described by means of the
\emph{spectral density} $J(\omega)$, which encodes the density of EM modes
and their coupling to the emitter. In this
Letter, we present a simple and easily implementable framework for
obtaining a few-mode quantum description of any given spectral
density. We start from a macroscopic QED treatment, which already provides the basis of quantization in
terms of a frequency continuum, and then construct a model system
consisting of a discrete number of \emph{interacting} modes
coupled to independent flat background baths, see \autoref{fig:1}.
Making use of Fano
diagonalization~\cite{Fano1961,Glutsch2002}, we obtain the general
form for the model spectral density, $\Jmod(\omega)$. This
can be fitted to any level of accuracy to $J(\omega)$, usually obtained by means of classical EM
calculations. We illustrate the power and validity of this
procedure in a hybrid structure comprising a plasmonic nanocavity
and a high-refractive-index microresonator. We show that our
approach enables accurate calculations of any far- and
near-field observables, proving that replacing the
full environment by our few-mode model does not lead to a loss of
information.

For a single emitter, the EM mode basis in macroscopic QED can be chosen such
that the Hamiltonian becomes ($\hbar=1$ here and in the
following)~\cite{Buhmann2008Casimir, Hummer2013, Rousseaux2016,
Sanchez-Barquilla2020}
\begin{multline}\label{eq:HMQED}
    H_{\mathrm{f}} = H_e +\! \int_0^{\infty} \mathrm{d}\omega\left[\omega a_\omega^{\dagger} a_\omega +
    \hat{\mu}_e g(\omega) (a_\omega + a_\omega^{\dagger}) \right],
\end{multline}
where $H_e$ is the bare emitter Hamiltonian and $\hat\mu_e$ is its
dipole operator, while $a_\omega$ are the bosonic annihilation
operators of the EM mode at frequency $\omega$. These fulfill
$[a_\omega,a_{\omega'}^\dagger] = \delta(\omega-\omega')$ and
correspond to the ``true modes'' in the nomenclature of
Refs.~\cite{Garraway1997Decay,*Garraway1997Nonperturbative,
Dalton2001}. The coupling between the emitter and the EM modes is
encoded by
\begin{equation}\label{eq:gMQED}
g(\omega)=\sqrt{\frac{\omega^2}{\pi \epsilon_0  c^2} \vec{n}\cdot
\text{Im}\{\mathbf{G}(\vec{r}_e,\vec{r}_e,\omega)\}\cdot \vec{n}},
\end{equation}
where $\vec{r}_e$ is the emitter position, $\vec{n}$ is the orientation of
its dipole moment (for simplicity, we assume that all relevant transitions are
oriented identically), and $\mathbf{G}(\vec{r},\vec{r}\,',\omega)$ is the
classical dyadic Green's function. This is directly related to the spectral
density of the environment, $J(\omega) = \mu^2 g(\omega)^2$ for a given
transition dipole moment $\mu$~\cite{Novotny2012}.

As discussed above, \autoref{eq:HMQED} requires the treatment of an EM
continuum. Without approximations, this is only possible with advanced and costly
computational techniques, such as tensor network
approaches~\cite{DelPino2018Dynamics,Zhao2020}. Our goal is thus to construct an
equivalent system that gives rise to
dynamical equations that can be solved easily. Our model environment (sketched
in \autoref{fig:1}) consists of $N$ interacting EM modes with ladder operators
$a_i$, $a_i^\dagger$, linearly coupled to the quantum emitter. Each of them is
also coupled to an independent Markovian (spectrally flat) background bath. The
resulting Hamiltonian is $\mathcal{H}= H_S + H_B$ with
\begin{subequations}\label{eq:Hmodel}
\begin{align}
    H_S &= H_e + \sum_{i,j=1}^{N} \omega_{ij} a^\dagger_{i} a_{j} + \hat{\mu}_e \sum_{i=1}^N g_i (a_i+a_i^\dagger),\\
    H_B &= \sum_{i=1}^{N} \int\left[\Omega b_{i,\Omega}^\dagger b_{i,\Omega} + \sqrt{\frac{\kappa_i}{2\pi}} (b_{i,\Omega}^\dagger a_i + b_{i,\Omega} a^\dagger_i)\right]
    \mathrm{d}\Omega.
\end{align}
\end{subequations}
$H_S$ is the system (emitter + discrete modes) Hamiltonian, where
the real symmetric matrix $\omega_{ij}$ describes the mode
energies and their interactions, and the real positive vector
$g_i$ describes their coupling to the emitter. The bath
Hamiltonian $H_B$ contains both the continuous bath modes,
described by the bosonic operators $b_{i,\Omega}$ and
$b_{i,\Omega}^\dagger$ at frequency $\Omega$, and the coupling
between the baths and the system, characterized by the rates
$\kappa_i$.

The power of our approach lies in the fact that the Hamiltonian above can be
analytically treated in two different ways: First, since the background baths
are completely flat, the Markov approximation is exact and the dynamics
described by $\mathcal{H}$ is identically reproduced, as proven
recently~\cite{Tamascelli2018}, by a Lindblad master equation
\begin{align}\label{eq:Lmodel}
    \dot\rho = -i [H_S,\rho] + \sum_i \kappa_i L_{a_i}[\rho],
\end{align}
where $\rho$ is the system density matrix and $L_O[\rho] = O\rho O^\dagger -
\frac12 \{O^\dagger O, \rho\}$ is a standard Lindblad dissipator. Secondly, the
linearly coupled system of $N$ interacting modes and continua can be
diagonalized by adapting Fano diagonalization for autoionizing states of atomic
systems~\cite{Fano1961}, which in this context is related to the theory of
quasimodes and
pseudomodes~\cite{Garraway1997Decay,*Garraway1997Nonperturbative,Dalton2001}.
This strategy allows us to obtain a simple, closed expression for
$\Jmod(\omega)$.

\begin{figure*}[t]
\includegraphics[width=0.95\linewidth]{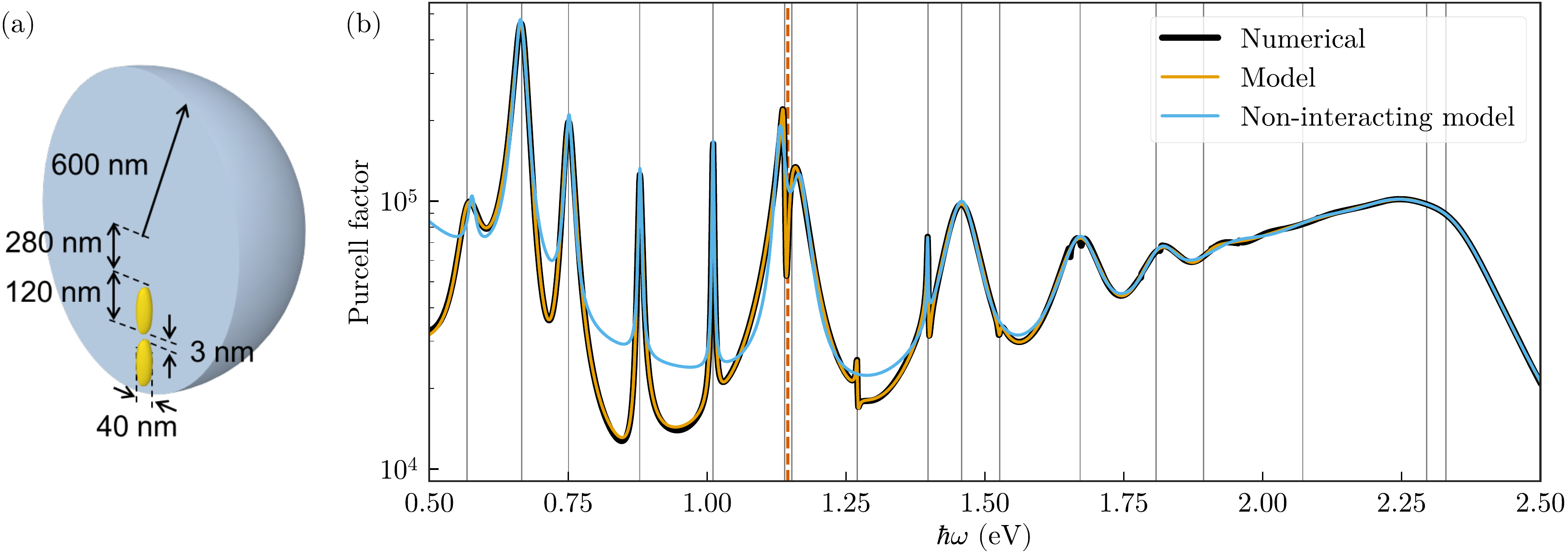}
\caption{(a) Sketch of the model system consisting of a silver
dimer nanoantenna embedded in a dielectric microsphere. (b)
Purcell factor $J(\omega)/J_0(\omega)$ for the system (thick black
line), for the fitted model described by \autoref{eq:Jmodel} with
19 modes (orange line), and for a model without interactions with
the same number of modes (light blue line). Thin gray lines
indicate the energy positions of the eigenstates of $\Heff$ for
the fitted interacting system. The red dashed line indicates the
frequency of the emitter used in \autoref{fig:3} and
\autoref{fig:4}.}\label{fig:2}
\end{figure*}

We first formally discretize the EM continua and rewrite \autoref{eq:Hmodel} as
\begin{equation}
    \mathcal{H} = H_e + \vec A^{\dagger T} \mathbf{H} \vec A + \hat{\mu}_e \vec M \cdot (\vec A + \vec
    A^\dagger), \label{eq:HFano}
\end{equation}
where $\vec A^T = (a_1, \ldots, a_N,
b_{1,\Omega_1},\ldots,b_{N,\Omega_n})$ collects the
annihilation operators of both modes and baths, the matrix
$\mathbf{H}$ describes their mutual couplings, and the vector
$\vec M^T = (g_1,\ldots,g_N,0,\ldots,0)$ collects their couplings
to the emitter (note that they vanish for the baths). Here, $N$ is
the number of discrete modes and associated continua, while $n$ is
the number of modes used to discretize each continuum.
Diagonalizing $\mathbf{H}$ gives the energies $\mathcal{E}_j$ and
eigenmodes $\vec{\psi}_j$ of the environment expressed in the
basis of the model system, and determines the spectral
density through $\Jmod(\omega) = \sum_j |\lambda_j|^2
\delta(\omega - \mathcal{E}_j)$, where $\lambda_j = \vec{M} \cdot
\vec{\psi}_j$. Using the Sokhotski-Plemelj formula, we can write
the spectral density in terms of the resolvent of
\autoref{eq:HFano} as ${\Jmod(\omega) = \tfrac{1}{\pi} {\rm
Im}\{\vec{M}^{\dagger T}(\mathbf{H}-\omega)^{-1}\vec{M}\}}$.
Applying a formalism developed for the calculation of the absorption spectrum in
atomic systems~\cite{Glutsch2002} and recovering the continuum limit for the
baths ($n\to\infty$), we finally obtain
\begin{equation}\label{eq:Jmodel}
\Jmod(\omega) = \frac{1}{\pi} \Im \left\{ \vec{g}^{\,T}
\frac{1}{\Heff - \omega} \vec g\right\},
\end{equation}
where $\vec{g}^{\,T} = (g_1, g_2, \ldots, g_N)$ is now an
$N$-element vector and the $N\times N$ matrix $\Heff$ has entries
$\Heff_{ij} = \omega_{ij} - \frac{i}{2} \kappa_i \delta_{ij}$.
Note that we have absorbed Lamb shifts of the modes due
to the coupling with the baths into the mode frequencies
$\omega_{ii}$ (as we have also implicitly done in \autoref{eq:Lmodel}). 
We note that as required for a spectral density, this form is non-negative,
i.e., $\Jmod(\omega)\geq 0$ for all $\omega$~\cite{supplemental}.

The last step in our approach consists in using
\autoref{eq:Jmodel} to fit $J(\omega)$ for a given EM environment,
which allows us to parameterize~\autoref{eq:Lmodel} for that
system. Although the number of unknowns in $\Jmod(\omega)$ is
relatively large ($N^2 + 2N$ real numbers for $\omega_{ij}$,
$\kappa_i$, and $g_i$), the fit procedure turns out to be stable
even for large numbers of modes (up to $N=19$ in the example
below). A similar procedure based on the fitting of the bath
correlation function has been reported as problematic
recently~\cite{Mascherpa2020}. Note that for non-interacting
modes, $\omega_{ij} = \omega_i
\delta_{ij}$, \autoref{eq:Jmodel} simplifies to a sum of
Lorentzians of the form $\Jmod(\omega) = \sum_i \tfrac{g_i^2}{\pi}
\frac{\kappa_i/2}{(\omega - \omega_i)^2 + \kappa_i^2/4}$. This
reproduces the well-known relation between Lorentzian spectral
densities and lossy modes~\cite{Imamoglu1994,Grynberg2010} that has been widely
used to quantize simpler EM environments such
as plasmonic cavities in the quasi-static
approximation~\cite{Gonzalez-Tudela2014, Delga2014,
Li2016Transformation, Varguet2019Non-hermitian}. The
introduction of interactions allows significantly more freedom in 
fitting $\Jmod(\omega)$, and in particular allows
for the representation of interference effects and the associated
Fano-like line shapes. In the supplemental
material~\cite{supplemental}, explicit expressions of
\autoref{eq:Jmodel} for 2-4 interacting modes are presented, as
well as their fitting to $J(\omega)$ in two recent
studies~\cite{Gurlek2018,Franke2019}. In the following, we
illustrate the power of our approach by considering a hybrid
nanophotonic structure, with a significantly more complex spectral
density.

\autoref{fig:2}(a) shows the system under study: a 600 nm radius
GaP~\cite{Cambiasso2017} microsphere ($\varepsilon_\mathrm{sph} =
9$) embedding two 120 nm long silver nanorods (with permittivity
taken from Ref.~\cite{Rakic1998}) separated by a 3 nm gap,
substantially displaced from the center of the
sphere. The microsphere by itself supports many
long-lived and delocalized Mie resonances, while the plasmonic
dimer sustains confined surface plasmons with strongly
sub-wavelength effective volumes~\cite{Baumberg2019}. The
interaction between these different modes leads to a complex EM
spectrum, shown in \autoref{fig:2}(b) through the
Purcell factor $P(\omega) = J(\omega) / J_0(\omega)$ for an
emitter located in the center of the nanorods. Here, $J_0(\omega)
= \frac{\omega^3 \mu^2}{6\pi^2\hbar\varepsilon_0 c^3}$ is the
spectral density in free space. The thick black line plots
classical EM simulations performed with the Maxwell's
equation solver implemented in COMSOL Multiphysics. This Purcell
factor, and the corresponding $J(\omega)$, presents a
large number of maxima, with several Fano-like profiles that
indicate interference effects as typical for
hybrid metallodielectric systems~\cite{Gurlek2018,Franke2019}.

In order to obtain a stable fit of $\Jmod(\omega)$ to $J(\omega)$ 
despite the large number of parameters, we consider the
non-interacting model (where $\omega_{ij} = \omega_i \delta_{ij}$)
as a starting point. This fit converges rapidly by using the
spectral positions, curvature, and amplitudes of the local maxima
in $J(\omega)$ to obtain initial guesses for the frequencies
$\omega_i$, loss rates $\kappa_i$, and coupling strengths $g_i$ of
the non-interacting model. This gives a good fit for many of the peaks 
[light blue line in \autoref{fig:2}(b)],
but strongly overestimates the background at lower frequencies,
and fails to reproduce the Fano-like asymmetric profiles that
originate from the hybridization between sphere and dimer modes.
Using the non-interacting fit as a starting point for
\autoref{eq:Jmodel} leads to rapid convergence, and as
shown by the orange line in \autoref{fig:2}(b), the resulting
spectrum is in almost perfect agreement with the numerical Purcell
factor over the full frequency range. Thus, we have constructed a
compact model with a relatively small number of quantized
interacting modes, each coupled to a flat background bath, that
fully represents the EM environment in the nanophotonic structure
in \autoref{fig:2}(a). Such an accurate fitting is not possible by
means of non-interacting modes, at least not without significantly
increasing the amount of modes considered. The thin grey lines in
\autoref{fig:2}(b) indicate the (real part) of the eigenenergies
of the matrix $\Heff$. These correspond to the complex resonance
positions (poles) of $J(\omega)$~\cite{Dalton2001}.

\begin{figure}[t]
\includegraphics[width=\linewidth]{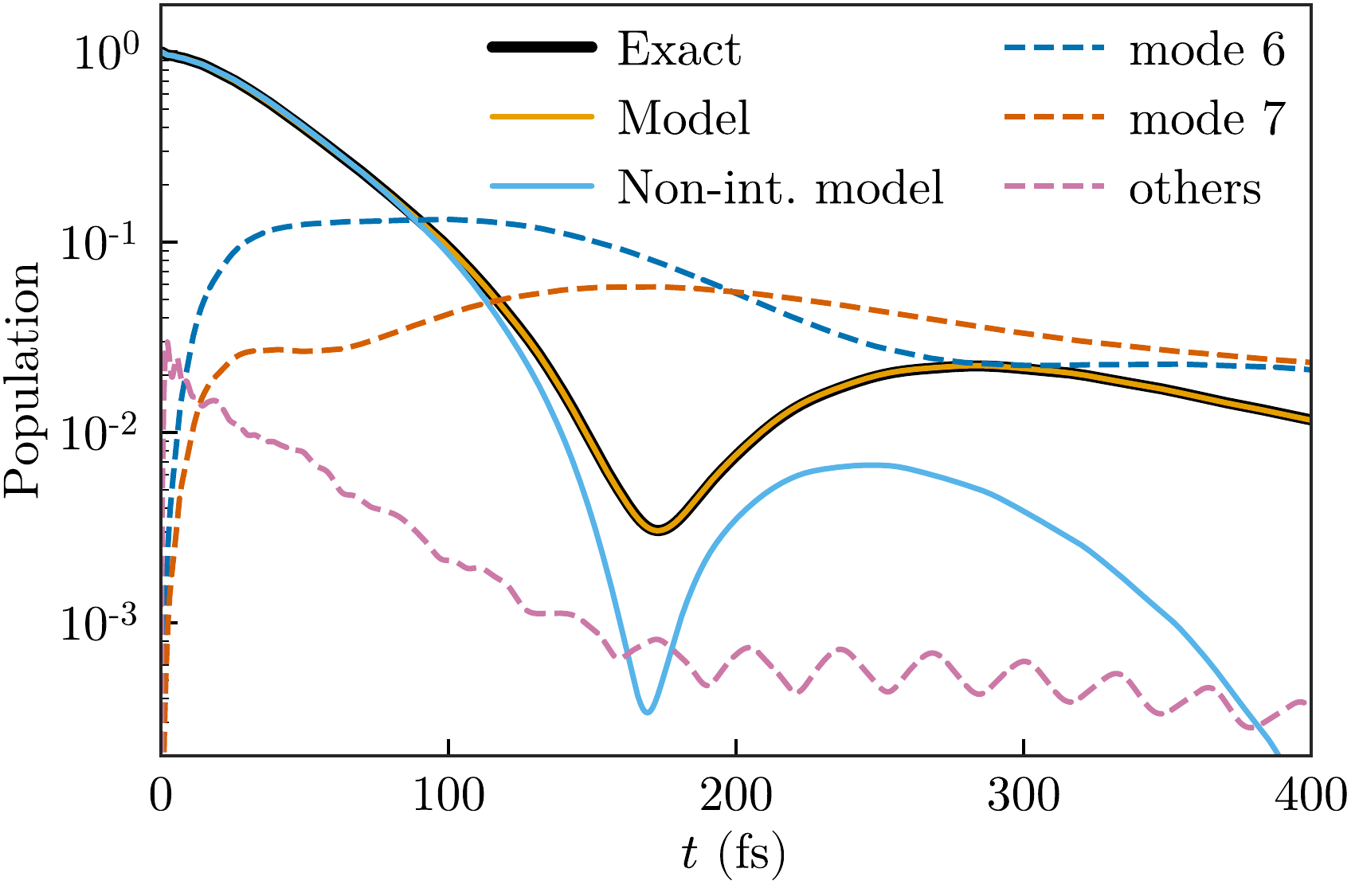}
\caption{Emitter and mode dynamics for the spontaneous emission
problem. Solid lines show the emitter excited-state population
$\langle \sigma^+\sigma^- \rangle(t)$ for the exact calculation
(black) and using the best-fit model with (orange) and without
(light blue) interactions. Dashed lines indicate the populations
$\langle \tilde{a}_\alpha^\dagger \tilde{a}_\alpha\rangle(t)$ of
modes $\alpha=6$ (blue) and $\alpha=7$ (red) (see text for
details), as well as the sum over all other mode populations
(magenta).}\label{fig:3}
\end{figure}

We next demonstrate that the model system indeed gives a
faithful representation of the EM environment, i.e., that the emitter dynamics with the model and with
the original spectral density are equivalent. To do so, we treat
the canonical spontaneous emission (Wigner-Weisskopf) problem for
a two-level emitter initially in its excited
state~\cite{Weisskopf1930}. We thus have $\hat{H}_e = \omega_{eg}
\sigma^+ \sigma^-$, $\hat{\mu}_e = \mu (\sigma^+ + \sigma^-)$,
where $\sigma^\pm$ are Pauli matrices. The emitter parameters are
chosen to represent InAs/InGaAs quantum dots~\cite{Eliseev2000}, with transition energy
$\hbar\omega_{eg} = 1.145$~eV, indicated by the dashed red line in
\autoref{fig:2}(b), and transition dipole moment $\mu =
0.55$~e\,nm.

\begin{figure*}
\includegraphics[width=0.95\linewidth]{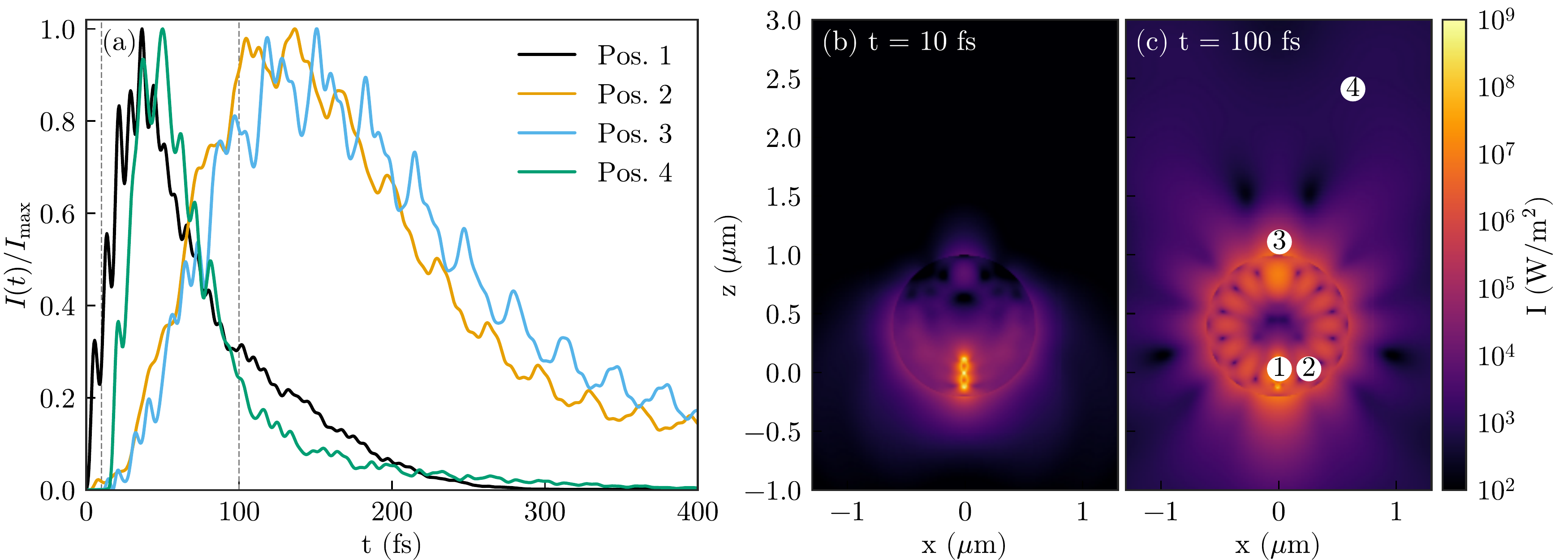}
\caption{Electric field intensity $\langle{\vec{E}^{(-)}(\vec{r})
\cdot \vec{E}^{(+)}(\vec{r})}\rangle$ for the initially excited
emitter as in \autoref{fig:3}. (a) Field intensity at four points,
shown by numbered white circles in panel (c), normalized to their
maximum values $I_{\max}$, given by $2.0\cdot 10^8$ W/m$^2$,
$1.8\cdot 10^6$ W/m$^2$, $1.1\cdot 10^6$ W/m$^2$, and $4.0\cdot
10^3$ W/m$^2$, respectively. (b) Field intensity distribution in
the $x$-$z$ plane at time $t=10$~fs, and (c) at
$t=100$~fs.}\label{fig:4}
\end{figure*}

Since in the Wigner-Weisskopf problem, there is at most one excitation in the
system (either in the emitter or in one of the EM modes), it can be solved
easily for arbitrary spectral densities~\cite{Grynberg2010}. The exact
excited-state population $\langle\sigma^+\sigma^-\rangle(t)$ obtained through
this approach is shown in \autoref{fig:3} (thick black line). The emitter
dynamics in the model Lindblad master equation~\autoref{eq:Lmodel} is obtained
using QuTiP~\cite{Johansson2012,*Johansson2013}. The excited-state population in
the full model (orange line) reproduces the exact results perfectly, while the
non-interacting model (light blue) fails to do so and shows significant
deviations from the correct result after about $100$~fs.

In order to gain further insight into the relevance and meaning of
the discrete modes obtained in our fit, we also show the populations of the modes $\tilde{a}_\alpha$
obtained by diagonalizing $\omega_{ij}$. Since $\omega_{ij}$ is a
real symmetric matrix, this corresponds to an orthogonal
transformation of the modes $a_i$, with $\tilde{a}_\alpha = \sum_i
V_{\alpha i} a_i$~\footnote{We note that applying this transformation to the master
equation, \autoref{eq:Lmodel}, gives rise to mixed Lindblad
dissipators, i.e., terms containing $\tilde{a}_\alpha$ and
$\tilde{a}_\beta^\dagger$ with $\alpha\neq\beta$, and associated
rates $\tilde{\kappa}_{\alpha\beta}$. We note for completeness
that the original fit could equally well have been performed in
this basis, which would correspond to a fully equivalent model
system of $N$ discrete modes and $N$ baths in which the modes do
not interact, but each mode is coupled to all the baths.}. The
dashed lines in \autoref{fig:3} show the mode populations of modes
$\tilde{a}_6$ and $\tilde{a}_7$, which we find to be the only
significantly populated modes for the
chosen emitter parameters. They are exactly the modes close to
resonance to the emitter frequency, $\tilde{\omega}_6 = 1.129$~eV
and $\tilde{\omega}_7 = 1.149$~eV (see \autoref{fig:2}). The
sum over all other mode populations, shown as a dashed magenta
line in \autoref{fig:3}, remains small during the whole
propagation. This demonstrates that our approach also allows the
identification of the relevant modes in the dynamics.

Finally, we show that although the model system is written in
terms of discrete lossy modes, it retains the full information
about the EM near and far field. The
electric field operator for the modes $a_\omega$ within the
formulation we use, \autoref{eq:HMQED}, can be written
as~\cite{Feist2020}
\begin{align}\label{eq:EMQED}
\vec{E}^{(+)}(\vec{r}) = \int_{0}^{\infty} \vec{\mathcal{E}}(\vec{r},\omega) a_\omega d\omega,
\end{align}
where the field mode profile is given by
\begin{align}\label{eq:E_mode_MQED}
\vec{\mathcal{E}}(\vec{r},\omega) = \frac{\hbar \omega^2}{\pi
\epsilon_0 c^2 g(\omega)} \Im \{ \mathbf{G}(\vec{r},
\vec{r}_e,\omega)\} \cdot \vec{n}.
\end{align}
The calculation then proceeds by solving the Heisenberg equations
of motion for the mode operators $a_\omega(t)$, which can be
formally integrated to yield
\begin{align}
    a_\omega(t) = a_\omega(0)e^{-i\omega t} - i g(\omega) \int_0^t \hat{\mu}_e(t') e^{-i \omega (t-t')} \mathrm{d}t'.
\end{align}
Inserting into \autoref{eq:EMQED} and defining the temporal kernel
\begin{align}
\vec{K}(\vec{r},\tau) &= \frac{\hbar}{\pi\epsilon_0 c^2}
\int_0^\infty \omega^2 \Im \{ \mathbf{G}(\vec{r},
\vec{r}_e,\omega) \} \cdot \vec{n}\, e^{i\omega\tau}
\mathrm{d}\omega
\end{align}
gives compact expressions for, e.g., the electric field
intensity
\begin{multline}\label{eq:I_field}
\langle{\vec{E}^{(-)} \cdot \vec{E}^{(+)}}\rangle = \int_0^t
\mathrm{d}t' \int_0^t \mathrm{d}t'' \langle\hat{\mu}_e(t')
\hat{\mu}_e(t'')\rangle \\ \vec{K}(\vec{r},t-t')
\vec{K}^*(\vec{r},t-t''),
\end{multline}
where, for simplicity, we have assumed that the field is
initially in the vacuum state.

\autoref{eq:I_field} enables the calculation of the field
intensity anywhere in space through the emitter correlation
functions, which can be easily obtained from \autoref{eq:Lmodel}.
This is displayed for the spontaneous emission case of \autoref{fig:3} 
in \autoref{fig:4}, with panel (a) showing the temporal dependence
of the electric field intensity at various points in space, with
locations indicated by numbered white circles in panel (c). The
intensities at each point are normalized to their maximum value,
which is given in the figure caption. Panels (b) and (c) show
snapshots of the field intensity profile in space at $t=10$~fs (b)
and $t=100$~fs (c). A movie showing the full field distribution
evolving in time is available in the supplemental
material~\cite{supplemental}. We note explicitly that in
this spontaneous emission problem, there is no coherent field,
$\langle \vec{E}^{(+)} + \vec{E}^{(-)}\rangle$, and it is thus
necessary to calculate the field intensity to observe the emission
dynamics. Interestingly, the dynamics at points (1), next to the
emitter, and (4) in the far-field (at a distance of $1.5~\mu$m
from the emitter), are quite similar, reaching their maximum value
within a few tens of femtoseconds and then decaying rapidly. In
contrast, points (2) and (3) inside and just outside the
dielectric sphere (but at some distance to the nanorod dimer) show
a much slower build-up and decay of the field intensity in time.
The comparison against \autoref{fig:3} reveals that the largest
contribution to the field intensity at positions 2 and 3 is given
by the hybrid modes 6 and 7, while the initial fast decay is due to the
contribution of other modes and leads to an intense initial pulse
radiated from the system, as seen at position 4.

To conclude, we have presented a simple and insightful procedure
to quantize the electromagnetic field in arbitrary nanophotonic
systems. Our approach works at the level of the spectral
density, calculated through the solution of Maxwell's equations.
This is fitted to a model spectral density, obtained through Fano
diagonalization and involving only a small number of lossy and
interacting electromagnetic modes. This makes it possible to
construct and parameterize a few-mode master equation accurately 
describing the interaction of a quantum emitter with the original
EM environment. We have illustrated the power and
validity of our ideas by calculating the spontaneous emission
population dynamics and near- and far-field intensity for
an emitter placed within a hybrid structure comprising a
dielectric microresonator and a plasmonic cavity. Our findings
offer a versatile and easily implementable framework for the
theoretical description of quantum nano-optical phenomena with
Dyadic Green's function calculations as the single input.

\begin{acknowledgments}
The authors thank Diego Mart\'in-Cano for interesting discussions
and sharing their data. This work has been funded by the European
Research Council through grant ERC-2016-StG-714870 and by the
Spanish Ministry for Science, Innovation, and Universities –
Agencia Estatal de Investigación through grants RTI2018-
099737-B-I00, PCI2018-093145 (through the QuantERA program of the
European Commission), and MDM-2014-0377 (through the María de
Maeztu program for Units of Excellence in R\&D). It was also
supported by a 2019 Leonardo Grant for Researchers and Cultural
Creators, BBVA Foundation. I.~M. thanks the nanophotonics group
for the warm hospitality during his stay at UAM, and acknowledges 
funding from the Coordena\c{c}\~ao de Aperfei\c{c}oamento de Pessoal 
de N\'ivel Superior - Brazil (Capes) through grant 88887.368031/2019-00.
\end{acknowledgments}

\bibliography{extranotes,references}

\end{document}